\documentclass[a4paper,10pt]{article}
\usepackage{graphicx}
\usepackage{epstopdf}
\usepackage{amsfonts,amsmath,amssymb}
\usepackage{hyperref}

\usepackage{epsfig,multicol,bbm}

\newcommand\fverb{\setbox\pippobox=\hbox\bgroup\verb}
\newcommand\fverbit{\egroup\item[\fbox{\unhbox\pippobox}]}

\newbox\pippobox
\begin{document}
\title{\bf Dark side of the universe in the  Stephani cosmology}
\author{ S. Sedigheh Hashemi,\,\, S. Jalalzadeh\thanks{Electronic address: s-jalalzadeh@sbu.ac.ir}\,\, and\,\, N. Riazi
\\
\small Department of Physics, Shahid Beheshti University, G.C., Evin, Tehran 19839,  Iran}
\maketitle
\begin{abstract}
We investigate the late time acceleration of the universe in the context of the Stephani model. This solution generalizes those of  Friedmann-Lemaitre-Robertson-Walker (FLRW)
in such a way that the spatial curvature is a function of of time. We show that the inhomogeneity of the models can lead to an accelerated evolution of the universe that is analogous to that obtained with FLRW models through a cosmological constant or any exotic component for matter.
\end{abstract}
\section{Introduction}\label{sec1}
Cosmological observations of the cosmic microwave background (CMB) and the large scale distribution of galaxies indicate
 that
the universe is homogeneous and isotropic on scales larger than about  100 megaparsecs \cite{1}.
The apparent  acceleration of  the expansion of the universe deduced from type  Ia  supernova
observations and the CMB, WMAP and Planck data
\cite{2}-\cite{4}, is
one of the most striking cosmological observations of recent times. In the context of
 FLRW models with matter and radiation energy components, the acceleration of our universe
 can not be explained and
would require either the presence of a cosmological constant or a new form of
matter
which does not clump and  dominates the late time evolution  with  a negative
pressure \cite{5, 6}.
However, since there is no explanation for the presence of a cosmological
constant of the appropriate value  and there is no natural candidate for
dark energy, it is tempting to look for alternative explanations by simply taking one
step back and noticing one of the fundamental assumptions of cosmology, homogeneity \cite{7}.
Recently, inhomogeneous models of the universe have become popular among
cosmologists \cite{8}.
 One of the inhomogeneous cosmological models
 which is
 a solution of Einstein equations with perfect fluid source and  conformally flat is  the
Stephani model  \cite{8}-\cite{10}. The matter in these solutions
has zero shear and rotation and moves with acceleration \cite{11}.  This solution
was obtained by  Stephani in 1967. His solution emerged as one of the spacetimes that can be embedded in a flat
five dimensional space  \cite{12}. This  universe and some of its subcases have been examined in many papers
(see e.g. \cite{8} and the references therein).
It has also been used in  stellar models  \cite{13}-\cite{17} and some generalizations to  FLRW \cite{18, 19}. Other papers have pointed out  their
 singularities   \cite{20}-\cite{24}
 and the thermodynamics
of their fluid source
\cite{25}-\cite{27}. Recently,   models with inhomogeneous pressure for
 testing the  astronomical data related to supernovae observations have been put forward \cite{28, 29}.
In these solutions, the energy density of matter, $\rho$, depends only  on cosmic time,
$t$, while pressure, $p$, depends on both  $t$ and $r$ which  means that the temperature varies with spatial position.
One of the objections to this model is their incompatibility with a linear barotropic equation of the form $p(t)=w_{0}\rho(t)$ with a constant $w_{0}$
\cite{27}. However, this incompatibility would not prohibit one to dismiss this solution since the  non-barotropic
equation of state can not be ruled out a priori.        
 The solution has
a curvature parameter which is time dependent and does not have any special form. Generally,
the  curvature parameter can be positive at one time and  negative at another time, thus this
spacetime is interesting for its topological dynamics among cosmologists. The evolution of
the model depends on some arbitrary functions, accordingly it can  be named as private universe \cite{30}.
In the work by
Stelmach and Jakacka  \cite{31}, they assume that the curvature parameter relates to scale factor
by $K(t)=\beta R(t)$  where  $\beta $ is a  constant.
They showed that with $\beta<0$
a cosmological model with dust source can lead to the accelerated expansion at later times of evolution.

In this paper, we use the energy balance  equation to obtain  the pressure and energy density. We
assume a non-barotropic equation of state to provide  a reasonable interpretation for the Stephani universe.
We  consider an  ansatz of the type $K(t)=\beta(\frac{R}{R_{0}})^n$ for studying  the model.
 We show that the field equation  is similar to the Cardassian model where the expansion of the universe is accelerated without postulating any exotic
 matter field \cite{34}.
 We find  that for some specific values of $n$ the model exhibits accelerated
expansion at later stages of evolution which is in agreement with the recent Planck data \cite{32}.

This paper is organized as follows:
In section two, the Stephani universe is studied as an alternative model
of the universe. The pressure as a function of spacetime coordinates is derived.  In section three, we calculate  observable quantities such as
the Hubble and the deceleration  parameters. We show that  in the considered model a negative deceleration parameter can be observed which is
in agreement with the recent observations  \cite{32}.
 We derive a general statement  for the age of the universe in the  model. We show that how positive accelerated expansion of the universe can appear  in the model without using any exotic matter.
In the last section, we will  draw some conclusions.

\section{Spherically symmetric Stephani universe}\label{sec 2}
The line element of the spherically symmetric Stephani universe in comoving coordinates is usually given in
 such a way as to emphasize its similarity to FLRW models \cite{8}. However, we use the following alternative
 coordinates
\cite{11} which is more appropriate for simplifying our calculations that we are going to do in the next section
 \begin{align}\label{eq1}
 ds^2=&-D^2dt^2+V^2\left[dr^2+f^2\left(d\theta^2+\sin^2\theta d\phi^2\right)\right],\nonumber\\
 D=&\frac{1+F^2\left(K-RK_{,R}\right)}{1+KF^2}, ~~~~V= \frac{R}{1+KF^2},
 \end{align}
where $K(t)$ is the curvature parameter, $R(t)$ is the scale factor
 and $K_{,R}=\frac{K_{,t}}{R_{,t}}=\frac{dK}{dR}$,
 and the functions $f(r), F(r)$ are defined by three possible  combinations
 \begin{align}\label{eq2}
 &f=r,\qquad\quad \, F=\frac{r}{2},\nonumber\\
 &f=\sin r,\quad ~~ F=\sin\left(\frac{r}{2}\right),\\
 &f=\sinh r,\quad F=\sinh\left(\frac{r}{2}\right).\nonumber
 \end{align}
The general transformation that relates the relations (\ref{eq2}) with the radial coordinate $r$  and
the radial coordinate of the
Stephani Universe, $ \tilde{r}$, is given by $r=\int \dfrac{d \tilde{r}}{1+k_{0}\tilde{r}^2/4}$
 where $k_{0}=0,\pm 1$. Also we assume that the energy-momentum tensor  is that of  a perfect fluid
 \begin{equation}\label{eq3}
 T^{\mu \nu}=\left(\rho+p\right)u^\mu u^\nu+pg^{\mu \nu},
 \end{equation}
where $u^\mu =\frac{1}{D}\delta^{\mu}_{t}$  is the fluid 4-velocity and $\rho, p$ are the
 mass-energy density and the pressure, respectively. The time coordinate in metric (\ref{eq1})
 has been selected such that the expansion scalar $\Theta=u^\mu_{~~;\mu}$
reads as
\begin{equation}\label{eq4}
\Theta=3\frac{R,{t}}{R}.
\end{equation}
The matter in this universe  has zero shear and rotation but moves with acceleration which is defined as
 $\dot{u}_{\mu} \equiv u_{\mu;\nu}u^{\nu}$,
whose value for the metric   (\ref{eq1}) is
\begin{equation}\label{eq5}
\dot{u}_{\mu}=\frac{D_{,r}}{D} \delta^r_{\mu}=-\frac{RK_{,R}f}{2\left[1+KF^2\right]\left[1+F^2\left(K-RK_{,R}\right)\right]}\delta^r_{\mu}.
\end{equation}
Hence the covariant form of the energy balance  $  u_{\mu}T^{\mu\nu}_{~~~;\nu}=0$  and the momentum balance
$h_{\mu \nu}T^{\mu \sigma}_{~~~;\sigma}=0$
takes the following forms, respectively
\begin{equation}\label{eq6}
  u_{\mu}T^{\mu\nu}_{~~~;\nu}=\dot{\rho}+\left(\rho+p\right)\Theta=0,
\end{equation}
and
\begin{equation}\label{eq8}
h_{\mu \nu}T^{\nu \sigma}_{~~~;\sigma}=h_{\mu}^{~~\nu}p_{,\nu}+(\rho+p)\dot{u}_{\mu}=0,
\end{equation}
where $\dot{\rho}=u^{\mu}\rho_{,\mu}$ is the proper time
derivative and $h^{\mu \nu}$ is the  projection tensor $h^{\mu \nu}=u^{\mu}u^{\nu}+g^{\mu \nu}$.

By inserting the line element  (\ref{eq1}) and the energy-momentum tensor (\ref{eq3}) into Einstein field equations the time-time component of
the field equations will be
\begin{equation}\label{eq7}
 \left(\frac{R_{,t}}{R}\right) ^2+\frac{\left(K+k_{0}\right)}{R^2}=\frac{8 \pi G}{3}\rho,
\end{equation}
where
\begin{equation}\label{piece}
k_{0}=\begin{cases}
0,&  $for$ \hspace{.3cm} f=r,  \hspace{.3cm} F=\dfrac{r}{2},\\
1,& $for$ \hspace{.3cm}  f=\sin(r),\hspace{.3cm} F=\sin\left(\dfrac{r}{2}\right)
,\\-1,& $for$ \hspace{.3cm} f=\sinh(r),\hspace{.3cm}  F=\sinh\left(\dfrac{r}{2}\right).
\end{cases}
\end{equation}
Eq.\;(\ref{eq7}) shows that $\rho$  is a function  of $t$ only.
We can obtain the form of pressure $p(r,t)$ from  (\ref{eq6}) as
\begin{equation}\label{eq9}
 p(r,t)=-\rho(t)- \frac{R}{3}\frac{\rho_{,t}}{R_{,t}}\left[\frac{1+KF^2}{1+F^2\left(K-RK_{,R}\right)}\right],
\end{equation}
and from the equation  of momentum balance (\ref{eq8}) we arrive at
\begin{equation}
 \acute{p}+\frac{fR^2\rho_{,R}K_{,R}}{6[1+(K-RK_{,R})F^2]^2}=0,
\end{equation}
where prime denotes  derivative with respect to $r$.

The Stephani universe has two arbitrary functions of time $(K(t), R(t))$
whose values are not prescribed \cite{11}.
We suppose that the curvature parameter
and    the matter energy density have the  following power law forms \cite{34}
\begin{equation}\label{eq10}
K\left(t\right)=\beta \left(\frac{R}{R_{0}}\right)^n,
\end{equation}
and
\begin{equation}\label{eq11}
\rho=\rho_{0}\left(\frac{R}{R_{0}}\right)^\alpha,
\end{equation}
we set $\alpha=-3(1+w_{0})$ where $w_{0}$ is the equation of state parameter which exists in FLRW models, $\beta$ and  $n$ are  constants, $R_{0}$ and $\rho_{0}$
are the present values of the scale factor and matter energy density, respectively.
A time derivative of the Eq.\;(\ref{eq7}) leads to the Raychaudhuri equation given by
\begin{equation}\label{eq67}
\frac{R_{,tt}}{R}=-\frac{4\pi G}{3}(1+3w_{0})\rho-\frac{\beta n}{2R_{0}^2}\left(\frac{R}{R_{0}}\right)^{n-2}.
\end{equation}
 Inserting
(\ref{eq10}) and (\ref{eq11})  into (\ref{eq9}) leads to
\begin{equation}\label{eq14}
p(r,t)=\left(-1+\frac{\left(1+w_{0}\right)(1+\beta (\frac{R}{R_{0}})^nF^2)}{1+\beta (\frac{R}{R_{0}})^n
(1-n)F^2}\right)\rho(t).
\end{equation}
Note that at the symmetry center $(r\simeq 0)$ for a matter-dominated universe $(w_{0}=0)$ pressure vanishes, but at large distances from the symmetry center
it will be negative, thus in the presence of a non-relativistic matter, due to a negative pressure at large distances, the expansion can be observed.
By comparing  Eq.\;(\ref{eq14}) with the general  equation of state with a non-constant parameter $w\equiv \dfrac{p}{\rho}$  we get
\begin{equation}
w(r,t)=\left(-1+\frac{\left(1+w_{0}\right)(1+\beta (\frac{R}{R_{0}})^nF^2)}{1+\beta (\frac{R}{R_{0}})^n
(1-n)F^2}\right),
\end{equation}
which can be regarded as the dynamical
equation of state parameter for the Stephani universe. For $\beta=0$ (switching off the inhomogeneities), we have $w=w_{0}$   and the  model reduces to the FLRW universe.

\section{Hubble and deceleration parameters}\label{sec 2a}
In order to study the cosmological parameters  which can be determined via observational data, it is important
to derive some of the  observational quantities in the Stephani universe.
The kinematics of the universe is  described by the Hubble parameter $H$ and the deceleration parameter $q$.
 For calculating these parameters we rewrite the metric (\ref{eq1}) in the comoving coordinate $[\tau, r, \theta, \phi]$
which reduces to
\begin{equation}
ds^2=-d\tau^2+V^2\left[dr^2+f^2\left(d\theta^2+\sin^2\theta d\phi^2\right)\right],
\end{equation}
where we have defined
\begin{equation}
D^2dt^2=d\tau^2.
\end{equation}
Now the definitions for $H $
 and $q$   will be
\begin{equation}
H=\frac{1}{V}\frac{\partial V}{\partial \tau},~~~~ q=-\frac{1}{H^2V}\frac{\partial ^2V}{\partial \tau^2}.
\end{equation}
Note that due to the inhomogeneity that occurs in this universe, the Hubble and deceleration parameters will no longer  be spatially constant. However, it is possible
to choose a time parameter in which  the spatial dependence of the Hubble and deceleration parameters can vanish \cite{31}.
Consequently,  the above definition for the Hubble parameter reduces to
\begin{equation}
H=\frac{1}{R}\frac{dR}{dt}.
\end{equation}
Also the deceleration parameter reads
\begin{equation}\label{eq18}
q\left(r,t\right)=-\frac{1+\beta \left(\frac{R}{R_{0}}\right)^nF^2}{1+\beta (1-n)\left(\frac{R}{R_{0}}\right)^n
F^2}\frac{R_{,tt}}{H^2 R}+\frac{n\beta \left(\frac{R}{R_{0}}\right)^nF^2}{1+\beta(1-n) \left(\frac{R}{R_{0}}\right)^n
F^2},
\end{equation}
which can be rewritten as
\begin{equation}\label{eq18a}
q\left(r,t\right)=\frac{\left(1+\beta\left(\frac{R}{R_{0}}\right)^nF^2\right)\tilde{q}+n\beta \left(\frac{R}{R_{0}}\right)^nF^2}{1+\beta (1-n)\left(\frac{R}{R_{0}}\right)^n
F^2},
\end{equation}
where
\begin{equation}
\tilde{q}\equiv-\dfrac{R_{,tt}}{H^2 R}=\frac{\frac{4\pi G }{3}\rho_{0}(1+3w_{0})\left(\frac{R}{R_{0}}\right)^{-3(1+w_{0})} +\frac{n\beta}{2R_{0}^2}\left(\frac{R}{R_{0}}\right)^{n-2}}{\frac{8\pi G}{3}\rho_{0}\left(\frac{R}{R_{0}}\right)^{-3(1+w_{0})}-\frac{k_{0}}{R_{0}^2}\left(\frac{R}{R_{0}}\right)^{-2}
 -\frac{\beta}{R_{0}^2}\left(\frac{R}{R_{0}}\right)^{n-2}}.
\end{equation}

With $n=1$ Eq.\;(\ref{eq18a}) reduces to the deceleration parameter obtained in the work by Stelmach and Jakacka \cite{30} in which they showed that
with a negative $\beta$ the deceleration parameter decreases with increasing distance to  the observed galaxy.

\begin{figure}[htp]
\centering\hspace{9.cm}\includegraphics[scale=0.56]{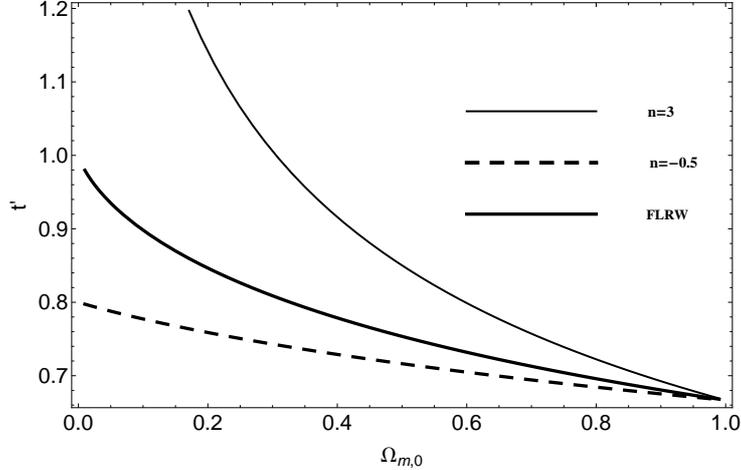}\caption{\label{9876} \small
 Lookback time $t\rq{}=tH_{0}$  in the spherically symmetric Stephani cosmological model for  some selected  values of $n$
 and in FLRW model  as a function of the energy density $\Omega_{m,0}$. }
\end{figure}

 The resemblance
 of the Eq.\;  (\ref{eq7}) with the first dynamical equation of the FLRW universe allows us to
 insert  Eqs.\;   (\ref{eq10}) and  (\ref{eq11}) into Eq.\;   (\ref{eq7})   to get an
 expression between $H, H_{0}, \Omega$ and $R$
\begin{equation}\label{eq23}
\left( \frac{H}{H_{0}}\right)^2=\Omega_{k_{0},0}\left(\frac{R}{R_{0}}\right)^{-2}+
\Omega_{0,0}\left(\frac{R}{R_{0}}\right)^{-3(1+w_{0})}+\left(1-\Omega_{0,0}-\Omega_{k_{0},0}\right)\left(\frac{R}{R_{0}}\right)^{n-2},
\end{equation}
where
\begin{equation}
\Omega_{0,0}=\frac{\rho_{0}}{\rho_{cr}}\mid_{t=t_{0}}, ~~~\Omega_{k_{0,0}}=\frac{-k_{0}}{R_{0}^2H_{0}^2}.
\end{equation}
Moreover by setting  $t=t_{0}$  in Eq.\;  (\ref{eq23}) the value of the constant
$\beta$ is given by
 \begin{equation}
 \beta=\left(\Omega_{0,0}+\Omega_{k_{0},0} -1 \right)R_{0}^2H_{0}^2.
 \end{equation}
Now we can find   the age of the universe  by integrating Eq.\;     (\ref{eq23})
  which yields
\begin{equation}\label{eq26}
t_{0}=\frac{1}{H_{0}}\int^{1}_{0}\frac{dx}{x\sqrt {\Omega_{0,0}x^{-3(1+w_{0})}+
 \Omega_{k_{0},0}x^{-2}+\left(1-\Omega_{0}-\Omega_{k_{0},0}\right)x^{n-2}      }},
\end{equation}
here $x=\dfrac{R}{R_{0}}$ is the scale  of the universe in units $R_{0}$. With  $n=1$  Eq.\;(\ref{eq26}) corresponds to the age of the universe  obtained
in the work by Stelmach and Jakacka \cite{31}.
For the present epoch we assume that the universe is only composed of dust $(w_{0}=0)$ and for simplicity in our
calculations we put $k_{0}=0$  as we do in flat  FLRW models. Accordingly,  Eqs.\; (\ref{eq18}), (\ref{eq23}) and (\ref{eq26}) reduce to
\begin{equation}
q(r,t)=\frac{1}{2}(\frac{1+\beta (\frac{R}{R_{0}})^nF^2}{1+\beta (1-n)(\frac{R}{R_{0}})^nF^2})    ( \frac{      \Omega_{m,0}+n(\Omega_{m,0}-1)   (\frac{R}{R_{0}})^{n+1}} {   \Omega_{m,0}-n(\Omega_{m,0}-1)  (\frac{R}{R_{0}})^{n+1}})
+\frac{n\beta (\frac{R}{R_{0}})^nF^2}{1+\beta (1-n)(\frac{R}{R_{0}})^nF^2},
\end{equation}
\begin{equation}\label{eq27}
 t_{0}=\frac{1}{H_{0}}\int^{1}_{0}\frac{dx}{x\sqrt {\Omega_{m,0}x^{-3}+
\left(1-\Omega_{m,0}\right)x^{n-2}      }},
\end{equation}
and
\begin{equation}\label{eq40}
\left( \frac{H}{H_{0}}\right)^2=\Omega_{m,0}\left(\frac{R}{R_{0}}\right)^{-3}+\left(1-\Omega_{m,0}\right)\left(\frac{R}{R_{0}}\right)^{n-2},
\end{equation}
respectively.
The lookback time $(t\rq{}=tH_{0})$ as a function of the energy density $\Omega_{m,0}$ in the
model for some specific values of $n$ and  in the FLRW models  are  presented in  Fig.\,\ref{9876}.   We realize that the age of the universe in the discussed model  for $n>2$ is
larger than in  FLRW models corresponding to the same values of the parameters $ H_{0}=67$ $\textmd{Km $Mpc^{-1}$$ s^{-1}$}$ and  $\Omega_{m}=0.31$ \cite{32}. As an example according to Eq.\;(\ref{eq27}) by setting $n=3$ the age of the universe will be
 $14.38$  ‎$\textmd{Gyr}$.

\begin{figure}[htp]
\centering
 \hspace{1.cm}\includegraphics[scale=0.56]{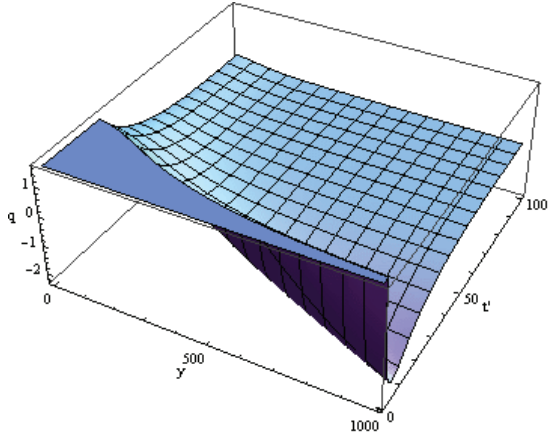}\caption{\label{nn} \small
The deceleration parameter in the spherically symmetric Stephani cosmological model as function of the dimensionless parameters  $y\equiv R_{0} H_{0}r$ and $t\rq{}=t_{0}H_{0}$ for $n=2.5$}
\end{figure}

At the late stage of the cosmological evolution when the scale factor is  large,  the first and the second terms in  Eq.\;(\ref{eq40}) can dominate and we can apply the following approximations.
From now on we assume that  $\mid\beta x^n F^2\mid\gg 1$. We consider Eq.\;(\ref{eq40}) in the late time epoch which can be solved for the scale factor to yield
\begin{equation}\label{eq41}
\frac{R}{R_{0}}\simeq\begin{cases}
\frac{1}{4}H_{0}^2(1-\Omega_{m,0})t^{2},& \hspace{.8cm}  n=1,\\
\exp[\sqrt{1-\Omega_{m,0}}H_{0}t],& \hspace{.8cm} n=2,\\
\left[H_{0}\left(1-\frac{n}{2}\right)\sqrt{1-\Omega_{m,0}}\right]^{\frac{2}{2-n} }t^{\frac{2}{2-n}},& \hspace{.8cm}  n>1,\quad  $or$ \quad-1<n<0,\\
(\frac{3}{2}H_{0}\sqrt{\Omega_{m,0}})^\frac{2}{3} t^{\frac{2}{3}},&\hspace{0.8cm} n<-1.
\end{cases}
\end{equation}
We should  point out that the case $n=2$ has a exponential form for the scale factor in the late time stage
which is the same as the vacuum  FLRW models with  cosmological constant.  However, as mentioned before,  the universe in this model  is only filled with dust which leads to the
above expressions. In the  case $n<-1$ we get a sort of dust dominated universe.
It follows from Eq.\;(\ref{eq41}) that  the deceleration parameter for the late time epoch can be derived for the above stages as
\begin{equation}\label{eq42}
q\left(r,t\right)\simeq\begin{cases}
\frac{1}{2}(\Omega_{m,0}-1)R_{0}^2H_{0}^2(\frac{R}{R_{0}}) F^2,& \hspace{.8cm}  n=1,\\
-1,& \hspace{.8cm} n=2,\\
\frac{n}{2(1-n)},& \hspace{.8cm}  n>1,\quad -1<n<0,\\
\frac{1}{1-n}(n+\frac{1}{2}),&\hspace{0.8cm} n<-1.
\end{cases}
\end{equation}
For all of the above values of $n$, the  deceleration parameter is negative, thus the accelerated expansion of the universe  is obtained without using any exotic matter or cosmological constant.
In  Fig.\,\ref{nn} the deceleration parameter is plotted for $n=2.5$ as a function of the dimensionless parameters $y\equiv R_{0} H_{0}r$ and $t\rq{}=t_{0}H_{0}$.
It is seen from Fig.\,\ref{nn} that the acceleration becomes larger as the distance to the observed object is increased. In other words, in the inhomogeneous universe acceleration of the expansion increases with the distance.

It should be also noted that Eqs.\;(\ref{eq27}) and (\ref{eq40}) can be rewritten as
\begin{equation}
H^2=\frac{8\pi G}{3}\rho_{m}+B\rho_{m}^{\gamma},
\end{equation}
where the constants $B$ and $\gamma$ are defined as
\begin{equation}
B=\frac{H_{0}^2(1-\Omega_{m,0})}{\rho^{\gamma}_{m,0}},\quad \gamma=\frac{2-n}{3}.
\end{equation}
The above equation is similar to the model proposed by Freese and Lewis \cite{32} which is an alternative model explaining  the accelerating universe. This model is called Cardassian model where the FLRW equation is
modified by the presence of the term $\rho^\gamma$. This seems to be interesting in the way that the expansion of the universe is accelerated automatically by the presence of the second term without suggesting any unknown form of exotic matter   \cite{33}.
\section{Conclusion}
In this paper we considered the inhomogeneous  Stephani universe characterized with a time dependent curvature index.
 Although we obtained results for general $w$ (equation of state parameter), but
 we focused our attention on the model with only dust as the fluid component of the universe. We found the age of the universe in the model which was remarkably larger than the corresponding age in the FLRW models without exotic matter.
We derived the deceleration parameter  which was dependent on the scale factor and radial coordinate, moreover we showed that the acceleration becomes larger while increasing the distance.
The fore-mentioned results were based on  the ansatz of the type $K(t)=\beta \left(\frac{R}{R_{0}}\right)^n$ for the spatial curvature parameter and we showed that  in the late time epoch of evolution,  by choosing the appropriate
values for $n$,  the power-law solution can be revived in the considered model.\\\\
\begin{center}
\textbf{Acknowledgment}
\end{center}
The authors would like to thank the anonymous referee for the enlightening elements.

\end{document}